\documentclass[a4paper,reqno]{amsart}
\usepackage{amssymb}
\usepackage{hyperref}
\usepackage{graphicx}
\usepackage{subcaption}
\usepackage{latexsym}
\usepackage{amsmath,amsthm,mathtools}
\usepackage{physics}
\usepackage{euscript}
\usepackage{graphics,color}
\usepackage[all]{xy}
\usepackage[margin=3cm]{geometry}
\usepackage{MnSymbol} 

\newcommand{\be}{\begin{equation}}
\newcommand{\id}{I}
\newcommand{\qbinom}[2]{\bigl[\begin{smallmatrix}
        #1\\#2
    \end{smallmatrix}\bigr]}
\newcommand{\ee}{\end{equation}}
\newcommand{\ba}{\begin{eqnarray}}
\newcommand{\ea}{\end{eqnarray}}
\newcommand{\baa}{\begin{eqnarray*}}
\newcommand{\eaa}{\end{eqnarray*}}
\newcommand{\bb}{\begin{thebibliography}}
\newcommand{\eb}{\end{thebibliography}}

\newcommand{\tends}{\rightarrow}

\newcounter{my}
\newcommand{\he}%
   {\stepcounter{equation}\setcounter{my}%
   {\value{equation}}\setcounter{equation}0%
   }%
\newcommand{\she}%
   {\setcounter{equation}{\value{my}}%
    }%

\newcommand{\Uqsu}{U_q(\mathfrak{su}(2))}

\newcommand{\qqbinom}[2]{\genfrac{[}{]}{0pt}{}{#1}{#2}}

\ExplSyntaxOn

\NewDocumentCommand{\pFq}{O{}mmmmm}
 {
  \group_begin:
  \keys_set:nn { hypergeometric } { #1 }
  \hypergeometric_print:nnnnn { #2 } { #3 } { #4 } { #5 } { #6 }
  \group_end:
 }
\NewDocumentCommand{\hypergeometricsetup}{m}
 {
  \keys_set:nn { hypergeometric } { #1 }
 }

\tl_new:N \l_hypergeometric_divider_tl
\tl_new:N \l_hypergeometric_left_tl
\tl_new:N \l_hypergeometric_right_tl

\keys_define:nn { hypergeometric }
 {
  symbol .tl_set:N = \l_hypergeometric_symbol_tl,
  symbol .initial:n = \phi,
  separator .tl_set:N = \l_hypergeometric_separator_tl,
  separator .initial:n = {},
  skip .tl_set:N = \l_hypergeometric_skip_tl,
  skip .initial:n = 8,
  divider .choice:,
  divider/semicolon .code:n = \tl_set:Nn \l_hypergeometric_divider_tl { \;; },
  divider/bar .code:n = \tl_set:Nn \l_hypergeometric_divider_tl { \;\middle|\; },
  divider .initial:n = bar,
  fences .choice:,
  fences/brack .code:n = 
   \tl_set:Nn \l_hypergeometric_left_tl {[}
   \tl_set:Nn \l_hypergeometric_right_tl {]},
  fences/parens .code:n = 
   \tl_set:Nn \l_hypergeometric_left_tl {(}
   \tl_set:Nn \l_hypergeometric_right_tl {)},
  fences .initial:n = brack,
 }

\cs_new_protected:Nn \hypergeometric_print:nnnnn
 {
  {} \sb {#1} \l_hypergeometric_symbol_tl \sb { #2 }
  \left\l_hypergeometric_left_tl
  \genfrac .. 
           {0pt} 
           {} 
           { \__hypergeometric_process:n { #3 } } 
           { \__hypergeometric_process:n { #4 } } 
  \l_hypergeometric_divider_tl
  #5
  \right\l_hypergeometric_right_tl
 }

\cs_new_protected:Nn \__hypergeometric_process:n
 {
  \clist_use:nn { #1 }
   {
    {\l_hypergeometric_separator_tl}
    \mspace { \l_hypergeometric_skip_tl mu }
   }
 }

\ExplSyntaxOff

\theoremstyle{definition}

\numberwithin{equation}{section}

\title{a $q$-version of the relation between the hypercube, the Krawtchouk chain and Dicke states}

\author{Pierre-Antoine Bernard}
\address{Centre de Recherches Mathématiques, Université de Montréal, C.P. 6128, Succursale Centre-ville, Montréal, QC H3C 3J7, Canada}
\email{pierre-antoine.bernard@umontreal.ca}

\author{Etienne Poliquin}
\address{Centre de Recherches Mathématiques, Université de Montréal, C.P. 6128, Succursale Centre-ville, Montréal, QC H3C 3J7, Canada}
\email{etienne.poliquin.1@umontreal.ca}

\author{Luc Vinet}
\address{IVADO and Centre de Recherches Mathématiques, Université de Montréal, C.P. 6128, Succursale Centre-ville, Montréal, QC H3C 3J7, Canada}
\email{luc.vinet@umontreal.ca}

\begin{document}

\begin{abstract}
It is shown how the spin chain based on the dual $q$-Krawtchouk polynomials is connected to a weighted hypercube through the use of $q$-Dicke states. The representation theoretic underpinnings based on the quantum algebra $\Uqsu$ are emphasized.
\end{abstract}

\maketitle

\section{Introduction}
When considering $XX$ spin networks or many-(free) fermion problems, one typically \cite{ crampe2019free,vinet2012construct} has to deal with the stable one-excitation subspace on which the restriction of the Hamiltonian is nothing else than the (weighted) adjacency matrix of the underlying graph. In the case where the nearest-neighbor couplings and magnetic fields are given by the recurrence coefficients of the Krawtchouk polynomials \cite{koekoek2010hypergeometric}, it is known that this matrix 
can be obtained as a projection of the restricted Hamiltonian of a hypercubic network with identical couplings \cite{bernard2018graph, bernard2024entanglement,christandl2004perfect}. In graph theoretical language, the one-dimensional structure is a weighted path that can be obtained as a quotient from the hypercube. This dimensional reduction is achieved with the help of the Dicke states that are totally symmetric multi-qubit state vectors. These Dicke states arise in many contexts such as quantum networking \cite{prevedel2009experimental} and QAOA \cite{farhi2014quantum} (see \cite{raveh2024q} for more references). As a matter of fact, these aforementioned observations prove to all be connected to the Lie algebra $\mathfrak{su}(2)$. Indeed, the Krawtchouk polynomials appear in the matrix elements of the $\mathfrak{su}(2)$ representations \cite{koornwinder1982krawtchouk, vinet2021note}, the hypercube is one of the graphs of the Hamming association scheme with $\mathfrak{su}(2)$ as its Terwilliger algebra \cite{bernard2023entanglement,go2002terwilliger} and the Dicke states span the highest dimensional irreducible $\mathfrak{su}(2)$ module contained in the regular representation of this algebra on the space with the vertex characteristic vectors of an hypercube as basis.

The question we address in this paper is whether a similar picture holds under $q$-deformation. In particular, is there some $q$-analog of the hypercube, together with a weighted adjacency matrix, that would project to the one-excitation restriction of the dynamics of a $XX$ spin chain based on a certain family of $q$-deformed Krawtchouk polynomials?

This will be answered in the affirmative. The leading concept will be that of the $q$-deformed or more simply $q$-Dicke states whose definition provided in \cite{li2015entanglement,raveh2024q} is rooted in the properties of the quantum algebra $\Uqsu$. This connection will also play a key role given the ubiquity of $\mathfrak{su}(2)$ in the $q=1$ story.

The paper is organized straightforwardly. So as to provide the appropriate background, the next section will review the connection between the hypercube and the Krawtchouk chain, bring up the role of the Dicke states and give details on the relevant $\mathfrak{su}(2)$ aspects that will support the $q$-extension. Next, in  Section \ref{s3}, it will be shown how the one-excitation subspace of an $XX$ spin network defined on a $q$-hypercube projects to the same sector for the $XX$ spin chain with couplings and magnetic fields given by the recurrence coefficients of the dual $q$-Krawtchouk polynomials. The adjacency matrix that fixes the weights of the hypercube and the dynamics of the network will be provided by the $\Uqsu$ twisted primitive element and its irreducible action on the $q$-Dicke states will be seen to yield the desired connection. The last section will contain concluding remarks.

\section{Krawtchouk chains and $N$-cubes}\label{s2}

\subsection{The Krawtcouk chain Hamiltonian and its one spin up restriction}

Consider the following Hamiltonian for an $XX$ spin chain with inhomogeneous nearest-neighbor couplings :
\begin{equation}
    \mathcal{H} = \frac{1}{2}\sum_{n=0}^{N-1} \sqrt{(n+1)(N-n)} (\sigma_n^x \sigma_{n+1}^x + \sigma_n^y \sigma_{n+1}^y).
\end{equation}
This Hamiltonian achieves perfect state transfer in that it enables the end-to-end transport of a qubit with perfect fidelity \cite{christandl2004perfect, vinet2012construct}. A necessary condition for that is the mirror symmetry ($n \rightarrow N-n$) \cite{kay2010perfect} that $\mathcal{H}$ is readily seen to possess. It is also observed that $\mathcal{H}$ commutes with the $z$-component of the total spin operator which ensures that the number of excitations is conserved. Let $\ket{n}$ refer to the state with a single spin up at position $n$, i.e.
\begin{equation}
    \ket{n} = \underbrace{\ket{0}\otimes\ket{0}\otimes \dots \ket{0}}_{n \text{ times}} \otimes \ket{1} \otimes \underbrace{\ket{0}\otimes\ket{0}\otimes \dots \ket{0}}_{N-n \text{ times}}.
\end{equation}
It is straightforward to verify that $\mathcal{H}$ has the following action on the states $\ket{n}$ :
\begin{equation}
    \mathcal{H}\ket{n} = \sqrt{(n+1)(N-n)} \ket{n+1} + \sqrt{n(N-n+1)} \ket{n-1}. \label{action H on n} 
\end{equation}
Denote by $H$ the restriction of $\mathcal{H}$ to the subspace spanned by these single spin up states $\ket{n}$. This matrix can be identified with the generator $j^x$ of $\mathfrak{su}(2)$ in an irreducible representation of spin $j = N/2$, where the basis is chosen as the normalized eigenvectors of $j^z$, $j^z\ket{j,m} = m \ket{j,m}$. Indeed, one finds that
\begin{equation}
    j^x \ket{j, m} = \frac{1}{2}\sqrt{(j-m)(j+m+1)} \ket{j, m+1} + \frac{1}{2}\sqrt{(j-m+1)(j+m)} \ket{j,  m-1},
\end{equation}
corresponds to equation \eqref{action H on n} upon the identifications $j^x\rightarrow H/2$, $j \rightarrow N/2$ and $m \rightarrow  n-N/2$. 
It follows that the eigenvalues of $H$ are readily obtained and that the associated equation may be written as
\begin{equation}
      H \ket{\omega_k} = \left(N - 2 k\right) \ket{\omega_k},\qquad  k \in \{0,1,\dots, N\}. \label{action H on omegak}
\end{equation}
Furthermore the eigenvectors $\ket{\omega _k} = \sum_{n=0}^{N} U_{nk} \ket{n}$ are determined by
the unitary matrix  $U_{nk} =\bra{n}\ket{\omega_k}$ that interchanges $j^x$ and $j^z$ in the irreducible spin $N/2$ representation, with entries known to be given in terms of Krawtchouk polynomials \cite{koornwinder1982krawtchouk}:
\begin{equation}
     \bra{n}\ket{\omega_k} = 2^{-N/2}\sqrt{\binom{N}{n}\binom{N}{k}} K_n(k; 1/2, N). \label{wavefn}
\end{equation}
This result is validated by recognizing that the wavefunctions $\phi_n(\omega_k) = \bra{n}\ket{\omega_k}$ satisfy the three-term recurrence relation
\begin{equation} \label{recurrence_relation_krawtchouk}
    (N-2k) \phi_n(\omega _k) = \sqrt{(n+1)(N-n)} \phi_{n+1}(\omega_k) + \sqrt{n(N-n+1)} \phi_{n-1}(\omega_k)
\end{equation}
in view of \eqref{action H on n} and \eqref{action H on omegak}. Allowing for the normalization factor in \eqref{wavefn}, it is seen that \eqref{recurrence_relation_krawtchouk} amounts to the three-term recurrence relation of the Krawtchouk polynomials \cite{koekoek2010hypergeometric,vinet2021note}.

\subsection{Connection with the hypercube}

We now recall how the Krawtchouk chain is obtained from a projection of a homogeneous model constructed on the hypercube graph $Q_N$. Let $V = \{0,1\}^N$ denote the set of bit sequences $x = (x_1, x_2, \dots, x_N)$ of length $N$, and let $\partial(x,y)$ stand for the Hamming distance between two sequences $x$ and $y$ which is defined as the number of positions where the sequences differ:
\begin{equation}
    \partial(x,y) = |\{ i \in \{1, 2, \dots, N\} \ |\ x_i \neq y_i\}|.
\end{equation}
The hypercube graph $Q_N$ has the set $V$ as its vertices, with edges connecting two sequences $x$ and $y$ if their Hamming distance $\partial(x,y)$ is $1$. To each sequence $x \in V$ is associated an orthonormalized vector $\ket{x} \in \mathbb{C}^{2^N}$. The adjacency matrix $A$ of $Q_N$ is the matrix whose entries are given by
\begin{equation}\label{def:A}
    \bra{x}A\ket{y} = \left\{
	\begin{array}{ll}
		1  & \mbox{if } \partial(x,y) =1  \\
		0 & \mbox{otherwise. }
	\end{array}
\right.
\end{equation}
Introduce now the column vectors  $\ket{D^N(n)}$ defined as the coherent sums of the vectors $\ket{x}$ corresponding to all the vertices at distance $n$ from the vertex

$\boldsymbol{0} = (0,0,\dots, 0)$:
\begin{equation}\label{def:col}
    \ket{D^N(n)} = \frac{1}{\sqrt{k_n}}\sum_{\substack{x \in V \\ \partial(x, \boldsymbol{0}) = n}} \ket{x}, \quad n \in \{0,1,...,N\},
\end{equation}
with $k_n =  \binom{N}{n}$, the number of sequences $x$ such that $\partial(x,\boldsymbol{0}) = n$. 
It is quite clear that $A$ will transform these column vectors among themselves and a simple combinatorial argument yields this action. It goes as follows \cite{bernard2024entanglement,christandl2004perfect}: pick a vector $\ket{x}$ entering in the sum defining $\ket{D^N(n)}$, there are $N-n$ vectors $\ket{y}$ composing $\ket{D^N(n+1)}$ to which $A$ can connect this $\ket{x}$; carrying then the sum over the $k_n$ vectors $\ket{x}$ at distance $n$ from $\boldsymbol{0}$ leads to 
$\bra{D^N(n+1)}A\ket{D^N(n) } = \sqrt{\frac{k_n}{k_{n+1}}}(N-n) = \sqrt{(n+1)(N-n)}$. Computing similarly $\bra{D^N(n-1)}A\ket{D^N(n) }$ one arrives at
\begin{equation}\label{eq:Aaction}
    A \ket{D^N(n)} = \sqrt{(n+1)(N-n)} \ket{D^N(n+1)} + \sqrt{n(N-n+1)} \ket{D^N(n-1)}.
\end{equation}
This equation matches with the action of the Hamiltonian of the Krawtchouk chain on the single spin-up states $\ket{n}$, with $A$ and $\ket{D^N(n)}$ playing the role of $H$ and $\ket{n}$ respectively. The Krawtchouk Hamiltonian $H$ can thus be obtained by projecting the adjacency matrix of an $N$-cube on the subspace of column states. 

The column vectors $\ket{D^N(n)}$ are actually the Dicke states, that is totally symmetric sums of $n$-qubit states. Their $\mathfrak{su}(2)$-based description will be given in the next subsection.

\subsection{The Dicke states and $\mathfrak{su}(2)$}

Association schemes \cite{bannai2021algebraic} can be viewed as sets of graphs with appropriate properties. In the case of the (binary) Hamming scheme \cite{bernard2023entanglement}, in addition to the hypercube, one has the additional graphs where it is the vertices at distance $2, \dots, N$ that are in turn connected. One can alternatively identify association schemes correspondingly with sets of adjacency (distance) matrices $A_i, \:  i=1, \dots, N$. It is also convenient to extend these sets of matrices by introducing the so-called dual adjacency matrices $A_i^*$ that we shall not define in general here. (We shall in the following use $A_1=A$ and $A_1^* = A^*$.) Roughly speaking then, the Terwilliger algebra of an association scheme is spanned by all the adjacency matrices and their duals \cite{terwilliger1992subconstituent}. In the case of the $N$-cube $Q_N$, the action of the dual adjacency matrix on a vector $\ket{x}$ amounts to measuring the distance  of $x$ to $\boldsymbol{0}$ :
\begin{equation}
    A^* \ket{x} = \left({N} - 2\partial(x, \boldsymbol{0})\right) \ket{x}.
\end{equation}

The Terwilliger algebra of the (binary Hamming) scheme involving the hypercube is isomorphic to $\mathfrak{su}(2)$ as can easily be argued with the following considerations. The complete graph with $\ell$ vertices $K_{\ell}$ has every pair of distinct vertices connected by a unique edge. It is easy to observe that the hypercube is the product of $N$ copies of $K_2$ : 
\begin{equation}
    Q_N = \underbrace{K_2 \times \dots \times K_2}_{N \; \text{times}}.
\end{equation}
Since the adjacency matrix of $K_2$ is the Pauli matrix $\sigma ^x$ and its dual $A^*$, the Pauli matrix $\sigma^z$, it follows that the adjacency matrix $A$ and dual adjacency matrix $A^*$ of $Q_N$ can be given as follows

\begin{equation}\label{A}
    A = \Delta^{(N-1)} (\sigma^x) = \sum_{i=1}^N \underbrace{I\otimes...\otimes I}_{i-1\text{ times}}\otimes\ \sigma_x\otimes \underbrace{I\otimes...\otimes I}_{N-i\text{ times}},
\end{equation}

\begin{equation}\label{dA}
    A^* = \Delta^{(N-1)} (\sigma^z) = \sum_{i=1}^N \underbrace{I\otimes...\otimes I}_{i-1\text{ times}}\otimes\ \sigma_z\otimes \underbrace{I\otimes...\otimes I}_{N-i\text{times}},
\end{equation}
where we used the coproduct $\Delta: \mathfrak{su}(2) \rightarrow \mathfrak{su}(2) \otimes \mathfrak{su}(2)$ defined by
\begin{equation}
    \Delta (j^a) = I \otimes j^a + j^a \otimes I, \qquad \text{for $a = x, y, z$},
\end{equation}
and the basis $\ket{x} = \ket{x_1}\otimes \dots \otimes \ket{x_n} \in \mathbb{C}^{2^N}$, with $\ket{0} = \binom{1}{0}$ and $\ket{1} = \binom{0}{1}$. We thus see that the Terwilliger algebra of the Hamming scheme (with $A$ and $A^*$ as generators) is isomorphic to $\mathfrak{su}(2)$.
Its representation on the $N$-fold tensor product of the spin $1/2$ representation is known to be reducible, with a spin $N/2$ component given by the so-called Dicke states :
\begin{equation}\label{def:dicke}
    \ket{D^N(n)} = \frac{1}{n!\sqrt{k_n}}\left(\Delta^{(N-1)}(\sigma^-)\right)^n \ket{\boldsymbol{0}}, \quad n \in \{0, 1, \dots, N\},
\end{equation}
where $\sigma^\pm = \frac{1}{2}(\sigma^x \pm i \sigma^y)$. It is straightforward to verify that these states correspond to the column vectors introduced in equation \eqref{def:col}, providing an interpretation of the projection of the hypercube onto the Krawtchouk chain in terms of the decomposition of $\mathbb{C}^{2^N}$ into irreducible $\mathfrak{su}(2)$ submodules. 

In the following, we introduce a $q$-deformation of the hypercube that maintains a similar connection to the $q$-analogs of Dicke states and Krawtchouk polynomials.

\section{$q$-Dicke states and $q$-hypercubes}\label{s3}

We now recall the definition of the quantum algebra $U_q(\mathfrak{su}(2))$ and of its coproduct. It has four generators denoted $e$, $f$, $k$ and $k^{-1}$ which satisfy the following defining relations,
\begin{gather}\label{uq(sl2)_def_1}
    k k^{-1}=k^{-1}k=\id,\qquad kek^{-1}=q^2e, \qquad kfk^{-1}=q^{-2}f \\ \label{uq(sl2)_def_2} [e,f]=\frac{k-k^{-1}}{q-q^{-1}}\equiv[h]_q,
\end{gather}
where $k\equiv q^{h}$ and $[x]_q\equiv\frac{q^x-q^{-x}}{q-q^{-1}}$. In the limit $q \tends 1$, \eqref{uq(sl2)_def_1} and \eqref{uq(sl2)_def_2} are readily seen to become the defining relations of $\mathfrak{su}(2)$ with $e \tends j^x + i j^y$, $f \tends j^x - i j^y$, $h \tends \frac{1}{2} j^z$, and $U_q(\mathfrak{su}(2))$ is thus seen to be the $q$-deformation of $U(\mathfrak{su}(2))$, the universal enveloping algebra of $\mathfrak{su}(2)$. Its fundamental representation is given in terms of Pauli matrices, with
\begin{equation}
    e\tends \sigma^+ \quad f\tends\sigma^-, \quad k\tends q^{\sigma^z}=
    \begin{pmatrix}
    q&0\\
    0&q^{-1}
\end{pmatrix},
\end{equation}
It is also endowed with a coproduct $\Delta_q:U_q(\mathfrak{su}(2)) \rightarrow U_q(\mathfrak{su}(2)) \otimes U_q(\mathfrak{su}(2)) $ defined as
\begin{gather}\label{eq:copq}
    \Delta_q(e)=e \otimes k^{-1/2} +k^{1/2} \otimes e, \qquad \Delta_q(f)=f \otimes k^{-1/2}+ k^{1/2} \otimes f \\ 
    \Delta_q(k) = k\otimes k, \quad \Delta_q(h) = h\otimes \id + \id \otimes h.
\end{gather}
The $q$-Dicke states are obtained by replacing the factorial coefficients and the coproduct of equations \eqref{def:dicke} by their $q$-analogs \cite{quasisymetric_qubit,raveh2024q}, 
\begin{equation}\label{def:dicke}
    \ket{D^N_q(n)} = \frac{1}{[n]_q!\sqrt{\qbinom{N}{n}_q}}\left(\Delta_q^{(N-1)}(\sigma^-)\right)^n \ket{\boldsymbol{0}}, \quad n \in \{0, 1, \dots, N\}.
\end{equation}
Similar to the $q=1$ case, the vectors $\ket{D^N_q(n)}$ can be written as a sum over sequences with $n$ 1s, with coefficients involving the \textit{inversion number} $\text{inv}(x)$ which is defined as the number of adjacent swaps needed to rearrange the binary sequence $x$ into the sequence with $N-n$ 0s followed by $n$ 1s. The $q$-Dicke states are expressed as \cite{quasisymetric_qubit,raveh2024q} :
\begin{equation}
\ket{D^N_q(n)} = \frac{1}{\sqrt{\qbinom{N}{n}_q}}\sum_{\substack{x \in V \\ \partial(x, \boldsymbol{0}) = n}} q^{\frac{n(N-n)}{2} - \text{inv}(x)}\ket{x}.
\end{equation}
As an example, we consider the sequences of four binary elements with two 1s. The associated inversion numbers and $q$-Dicke state are respectively
\begin{equation}
    \begin{pmatrix}
        \text{inv}(1100)\\\text{inv}(1010)\\\text{inv}(1001)\\\text{inv}(0110)\\\text{inv}(0101)\\\text{inv}(0011)
    \end{pmatrix} = \begin{pmatrix}
        4\\3\\2\\2\\1\\0
    \end{pmatrix},
\end{equation}
and
\begin{equation}
    \ket{D_q^4(2)}=\frac{1}{\sqrt{|q|^{-4}+|q|^{-2}+2+|q|^2+|q|^4}}\left(q^{-2}\ket{1100}+q^{-1}\ket{1010}+\ket{1001}+\ket{0110}+q\ket{0101}+q^{2}\ket{0011}\right).
\end{equation}

Just as the Dicke states correspond to the column vectors of the hypercube, one can introduce a weighted $N$-cube whose column vectors are the $q$-analogs $\ket{D^N_q(n)}$ of Dicke states. Consider the matrices $X^\pm$ and $K$, which are obtained by applying $N-1$ times the coproduct $\Delta_q$ to the fundamental representation of  $e$, $f$ and $k$, i.e.
\begin{equation}
    X^\pm = \Delta_q^{(N-1)}(\sigma^\pm), \quad K = \Delta_q^{(N-1)}(q^{\sigma^z}).
\end{equation}
The $q$-deformation of the adjacency matrix $A$ of the hypercube is then defined as,

\begin{equation} \label{A_q_def}
    A_q = (\sqrt{q} X^- + \frac{1}{\sqrt{q}}X^+)K^{-1/2} = \sum_{i=1}^N \underbrace{I\otimes...\otimes  I}_{i-1\text{ times}}\otimes\ \sigma_x\otimes \underbrace{ q^{-\sigma_z} \otimes...\otimes q^{-\sigma_z}}_{N-i\text{ times}}.
\end{equation}

\begin{figure}
\begin{subfigure}{.5\textwidth}
  \centering
  \includegraphics[width=.8\linewidth]{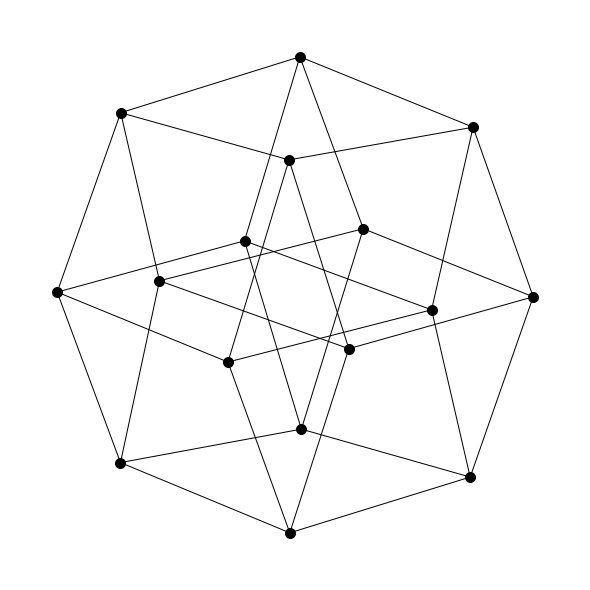}
  \caption*{$N = 4$, $q = 1$}
  \label{fig:sfig1}
\end{subfigure}%
\begin{subfigure}{.5\textwidth}
  \centering
  \includegraphics[width=.8\linewidth]{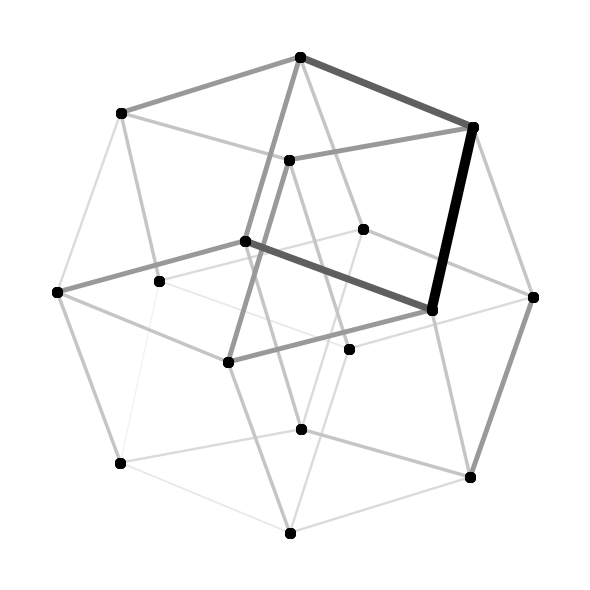}
  \caption*{$N = 4$, $ q = 0.7$}
  \label{fig:sfig2}
\end{subfigure}
\caption{$q$-deformed hypercubes. The edge color and width represent the amplitude of the weights.}
\label{fig:cube}
\end{figure}

This matrix corresponds to $Y = \left(\sqrt{q} f + \frac{1}{\sqrt{q}} e\right)k ^{-1/2}$ , a so-called twisted primitive element \cite{Koelink,koornwinder1993q}, in the $N$-fold tensor product of the fundamental representation. Specifically, $Y$ satisfies the condition $\Delta_q(Y) = Y \otimes K^{-1} + I \otimes Y$ and defines a co-ideal subalgebra. Furthermore, it is straightforward to see that picking this $A_q$ to describe a $q$-hypercube corresponds to the definition \eqref{def:A} of the hypercube, with additional weights on the edges that depend on the position $i$ where the two sequences differ :
\begin{equation}
    \bra{x}A_q\ket{y}=\left\{ \begin{array}{cc}
          q^{i-N + 2\sum_{j =i+1}^{N} x_j }& \text{if } \partial(x,y) = 1, \ x_{i} \neq y_{i} \\
        0 & \text{otherwise.} 
    \end{array}\right.
\end{equation}
The cases $N = 4$ with $q = 1$ and $q = 0.7$ are illustrated in Figure \ref{fig:cube} with the edge color and width representing the amplitude of the weights.

According to standard representation theory of $U_q(\mathfrak{su}(2))$, the $q$-Dicke states form an irreducible submodule of dimension $N+1$ within the module obtained from the tensor product of $N$ copies of the two-dimensional representation. Consequently, the weighted adjacency matrix $A_q$ of the $q$-deformed hypercube has a closed action on these column states.  This action is derived similarly to the action of $A$ on Dicke state, given in equation \eqref{eq:Aaction}. However, in the 
$q$-deformed case, additional identities related to the inversion number 
$\text{inv}(x)$ are required. Specifically, when 
$x$ and 
$y$ are sequences containing 
$n$ and 
$n+1$ 1s respectively and differing only at position 
$i$, then
\begin{equation}\label{eq:form1}
    \text{inv}(x) - \text{inv}(y) = n+ i - N.
\end{equation}
Moreover, a well-known formula for the summation of the inversion numbers over sequences at a fixed Hamming distance from $\boldsymbol{0}$, namely \cite{Stanley_2011}
\begin{equation}\label{eq:form2}
    \sum_{\substack{x \in V \\ \partial(x,\boldsymbol{0}) = n}} q^{n(N-n) - 2\text{inv}(x)} =   \qqbinom{N}{n}_q
\end{equation} proves handy.
The coefficients $\bra{D_q^N(n+1)}A_q\ket{D_q^N(n)}$ and $\bra{D_q^N(n-1)}A_q\ket{D_q^N(n)}$ are then computed using \eqref{eq:form1} and \eqref{eq:form2} with straightforward combinatorial reasoning and yield the following action of $A_q$ on the $q$-Dicke states
\begin{gather}\label{eq:projAq}
    A_q \ket{D^N_q(n)} =q^{n-\frac{N}{2}}\left(\sqrt{q[n+1]_q[N-n]_q\,} \ket{D^N_q(n+1)} +\sqrt{q^{-1}[n]_q[N-n+1]_q\,} \ket{D^N_q(n-1)} \right).
\end{gather}

Analogous to how the hypercube projects onto the single-particle Hamiltonian of the Krawtchouk spin chain, the restriction of $A_q$ to the span of the $q$-Dicke states yields a projection onto the single-particle subspace of the following Hamiltonian
\begin{equation}
    \mathcal{H}_q = \frac{1}{2}\sum_{n=0}^{N-1} q^{n- \frac{N}{2}}\sqrt{q[n+1]_q[N-n]_q} \ (\sigma_n^x \sigma_{n+1}^x + \sigma_n^y \sigma_{n+1}^y),
\end{equation}
Denote $H_q$ the matrix corresponding to the restriction of 
$\mathcal{H}_q$ to the single spin-up state. The action of $H_q$ on the states $\ket{n}$ is found to coincide with equation \eqref{eq:projAq}, establishing the correspondence: $H_q  \leftrightarrow A_q$ and $\ket{n} \leftrightarrow \ket{D_q^N(n)}$. It is further observed that the couplings in $H_q$ are related to the recurrence coefficients of the dual $q$-Krawtchouk polynomials, and that the one-particle wavefunctions are expressed in terms of these polynomials. Explicitly, the eigenstates $\ket{\omega_k}$ of $H_q$ are given by 
\hypergeometricsetup{symbol = \phi}
\begin{equation}
    H_q\ket{\omega_k}_q =[2k-N]_q\ket{\omega_k}_q, \quad  \ket{\omega_k}_q =\sum_{n=0}^N \hat{K}_n(\lambda(k);-1,N|q^2)\ket{n}, \quad k\in\{0,1,...N\},
\end{equation}
 with the wavefunction $\braket{n}{\omega_k}=\hat{K}_n(\lambda(k);-1,N|q^2)$ expressed as follows in terms of dual $q$-Krawtchouk polynomials $K(\lambda(x);c,N|q^2)$ \cite{koekoek2010hypergeometric} : 
\begin{gather} \label{normalized_dual_krawtchouk}
   \hat{K}_n(\lambda(x);c,\ell|q)\equiv \sqrt{\frac{(cq^{-\ell},q^{-\ell},q)_x (1-cq^{2x-\ell})c^{-x}q^{x(2\ell-x)}(q^{-\ell};q)_n}{(q,cq;q)_x(1-cq^{-\ell})(c^{-1};q)_\ell(q;q)_n(cq^{-\ell})^n}}K(\lambda(x);c,\ell|q), \\ K_n(\lambda(x);c,l|q)=\pFq{3}{2}{q^{-n},q^{-x},cq^{x-l}}{q^{-l},0}{q;q}, \quad c<0, \quad \lambda(x)=q^{-x}+cq^{x-\ell},
\end{gather}
where $(a;q)_n\equiv(1-a)(1-aq)\ \hdots \ (1-aq^{n-1})$ and $ (a_1,...,a_r;q)_n = (a_1;q)_n\ \hdots \ (a_r;q)_n$ are the $q$-Pochammer symbol. The role of the Krawtchouk polynomials in the diagonalization of the adjacency matrix of the hypercube, as detailed in Section \ref{s2}, thus carries over to the $q$-deformed case.

\section{Outlook}\label{s4}
We have introduced a $q$-analog of the hypercube as a weighted graph, and could hence extend the connections between the Krawtchouk spin chain, Dicke states, and the 
$N$-cube to their $q$-deformed counterparts. Specifically, we demonstrated that restricting the adjacency matrix $A_q$ of the weighted hypercube to the subspace spanned by $q$-Dicke states yields the Hamiltonian of a dual $q$-Krawtchouk spin chain acting on its stable one-excitation subspace.

It is important to note that other definitions of  $q$-deformed hypercubes exist in the literature, often involving the notion of a finite field  $\mathbb{F}_q$, where $q$ is a prime power. One such example is the subspace lattice $L_N(q)$ \cite{terwilliger1990incidence,watanabe2017algebra}, which is defined on the set of subspaces of a vector space of dimension $N$ over the field $\mathbb{F}_q$. This approach is motivated by the fact that $n$-subspaces of an $N$-dimensional vector space over a finite field are typically regarded as the combinatorial $q$-analog of $n$-subsets of a set with $N$ elements. Another example of $q$-analogs of hypercubes is found in the dual polar graphs from the theory of association schemes \cite{DRG, stanton1980some}. These graphs are defined on the set of maximal isotropic subspaces of a vector space over $\mathbb{F}_q$ equipped with a non-degenerate form. They are considered to be $q$-analogs of hypercubes since they are also distance-regular and give rise to a $P$-polynomial association scheme related to dual $q$-Krawtchouk polynomials.

It would be of genuine interest to investigate the connections between the physically motivated $q$-analog of the hypercube introduced here and these other lattices, all of which are linked to $U_q(\mathfrak{su}(2)) $ and dual $q$-Krawtchouk polynomials 
\cite{bernard2022terwilliger,terwilliger2023aq,watanabe2017algebra}. We plan on examining this through graph quotient in a future work.

\section*{Acknowledgments}

 The authors wish to thank Rafael Nepomechie for sharing with us his knowledge and insights about Dicke states and their generalisations. PAB holds an Alexander-Graham-Bell scholarship from the Natural Sciences and Engineering Research Council of Canada (NSERC). EP held an NSERC Undergraduate Student Research Award (USRA) while this research was conducted. The work of LV is funded in part through a NSERC discovery grant.

\end{document}